\documentclass{article}

\usepackage{arxiv}
\usepackage[utf8]{inputenc} 
\usepackage[T1]{fontenc}    
\usepackage{hyperref}       
\usepackage{url}            
\usepackage{amsfonts}       
\usepackage{nicefrac}       
\usepackage{microtype}      
\usepackage{lipsum}
\usepackage{fancyhdr}       
\usepackage{graphicx}       
\graphicspath{{media/}}     

\usepackage{caption}
\usepackage{subcaption}
\usepackage{amsmath}
\usepackage{amssymb}
\usepackage[ruled,vlined]{algorithm2e}
\usepackage{graphicx}
\usepackage{float}
\include{pythonlisting}
\usepackage{multirow}
\usepackage{makecell}
\usepackage{rotating}
\usepackage{booktabs}
\usepackage{xcolor}
\usepackage{natbib}

\usepackage{pgfplots}
\usepgfplotslibrary{fillbetween}
\pgfplotsset{width=10cm,compat=1.9}

\usepackage{algcompatible}

\algnewcommand\algorithmicdata{\textbf{Data:}}
\algnewcommand\DATA{\item[\algorithmicdata]}

\algnewcommand\algorithmicparameters{\textbf{Parameters:}}
\algnewcommand\PARAMETERS{\item[\algorithmicparameters]}

\algnewcommand\algorithmicbegin{\textbf{begin}}
\algnewcommand\BEGIN{\item[\algorithmicbegin]}

\algnewcommand\algorithmicEND{\textbf{end}}
\algnewcommand\END{\item[\algorithmicEND]}

\usepackage{amsthm}

\theoremstyle{plain}

\theoremstyle{definition}

\usepackage{xcolor}
\usepackage{amsmath}

\DeclareMathOperator*{\argmin}{arg\,min}
\newtheorem{prop}{Proposition}
\counterwithin{prop}{subsection}

\title{\textit{There is an elephant in the room}: Towards a critique on the use of fairness in biometrics
}

\author{
  Ana Valdivia \\
  King's College London (KCL) \\
  London, UK\\
  \texttt{ana.valdivia@kcl.ac.uk} \\
\AND
  Júlia Corbera-Serrajòrdia \\
  King's College London (KCL) \\
  London, UK\\
  \texttt{julia.corbera\_serrajordia@kcl.ac.uk}\\
  \And
  Aneta Swianiewicz \\
  King's College London (KCL) \\
  London, UK\\
  \texttt{aneta.swianiewicz@kcl.ac.uk}\\
}

\begin{document}
\maketitle

\begin{abstract}
In 2019, the UK's Immigration and Asylum Chamber of the Upper Tribunal dismissed an asylum appeal basing the decision on the output of a biometric system, alongside other discrepancies. The fingerprints of the asylum seeker were found in the EU's asylum fingerprint database and registered in 2016, which contradicted the appellant's account that he left Iraq in 2017. The Tribunal found this biometric evidence `unequivocal' and denied the asylum claim. Nowadays, the proliferation of biometric systems in our societies is shaping public debates around its political, social and ethical implications. Yet whilst concerns towards the racialised use of this technology for migration control or law enforcement have been on the rise, investment in the biometrics industry and innovation is increasing considerably. Moreover, fairness has also been recently adopted by biometrics to analyse demographic bias of biometric systems. Thousands of studies have been recently published to mitigate bias and discrimination on facial recognition, fingerprints or finger veins, among others, even suggesting a lack of bias on these systems. However, algorithmic fairness cannot distribute justice in scenarios which are broken or intended purpose is to discriminate, such as biometrics deployed at the border.

In this paper, we offer a critical reading of recent debates about biometric fairness and show its limitations drawing on research in fairness in machine learning and critical border studies. Building on previous fairness demonstrations, we prove that biometric fairness criteria are mathematically mutually exclusive. Then, the paper moves on illustrating empirically that a fair biometric system is not possible by reproducing experiments from previous works. Finally, we discuss the politics of fairness in biometrics systems by situating the debate around biometrics used at the border. We claim that bias and error rates have different impact on citizens and asylum seekers. Fairness has overshadowed the \textit{elephant in the room} of biometrics, focusing on the demographic biases and ethical discourses of these algorithms rather than examine how these systems reproduce historical and political injustices in these contexts.
\end{abstract}

  



\maketitle

\section{Introduction}
Biometric systems are being designed and implemented by public and private organisations for law enforcement, migration control and security purposes. This technology is used to identify or verify the unique identification of a person through physical, physiological or behavioural characteristics, such as fingerprint, face, voice, gait or finger vein of human bodies. Individual bodily characteristics are transformed into biometric data and used by the authorities as the \textit{truthful} identity of an individual. Biometric systems are part of our daily lives, we use them to unlock our mobile phones or to ‘efficiently’ cross borders at airports. However, this technology has different consequences depending on a subject's citizenship status. For instance, according to the  Dublin Regulation, the EU is extracting and storing fingerprints of asylum seekers in large-scale databases to control border crossings and determine the country responsible of the asylum petition~\cite{queiroz2019impact, broeders2007new, van1999illegal}. In other words, biometrics are implemented to immobilise, control and obstruct migrants' movements within Member States through fingerprints identification~\cite{scheel2013autonomy, tazzioli2019making}. The World Food Programme (WFP) in partnership with the United Nations High Commissioner for Refugees (UNHCR) implemented iris recognition to register migrants and provide cash assistance in refugee camps~\cite{aloudat2016implications}. The use of biometrics in these scenarios has been largely criticised by academics, activists and human \& digital rights organizations who have argued that `undermines democracy, freedom and justice'~\cite{edri2020fr, queiroz2019impact}. In the specific context of migration, biometric systems are used to illegalise freedom of movement through Dublin Regulation, infringing a fundamental right. Yet fairness has plunged into biometrics as a new research area which aims at debiasing these systems, ignoring the historical, political and social context in which this technology is embedded. 

This paper starts by questioning the ‘emergent challenge’ of fairness in biometrics. Since the publication of \textit{Gender Shades}~\cite{buolamwini2018gender} in 2018, the field of biometrics has experienced a surge in studies on bias and disparate impact in algorithmic systems such as facial recognition, fingerprints, finger veins, iris recognition, among others~\cite{drozdowski2020demographic}. By examining the recent literature on fairness in biometrics, we observe a lack of engagement with the state-of-the-art of fairness. Notably, a recent work in finger vein recognition systems have suggested a lack of bias for the tested biometric algorithms~\cite{drozdowski2021demographic}. Although finger vein systems have not been implemented for migration control, we propose to empirically demonstrate the impossibility of fairness examining the systems used in this work. Rather than evaluating fairness using at least one of the multiple definitions proposed in recent years by fairness scholars~\cite{hutchinson2019unfairness, verma2018fairness, dunkelau2019fairness, barocas-hardt-narayanan}, the authors analysed algorithmic discrimination by analysing statistical differences on descriptive statistics of the score distributions based on gender, age, fingers and hands. Although this might seem sensible to evaluate demographic biases, we outline serious limitations. First, the decision threshold plays a key role in the assessment of fairness in biometrics systems~\cite{de2020fairness}. Second, intersectional demographic evaluation of gender and age must be assessed. Third, it has been proved that fairness definitions are mutually exclusive, so the lack of bias is technically impossible in any algorithmic system~\cite{garg2020fairness, chouldechova2017fair,  zhao2019inherent, kleinberg2018inherent}. In order to show these flaws, this paper proceeds by developing a theoretical framework, translating fairness definitions into biometrics. Building on previous works in fairness and machine learning, we theoretically proved the impossibility of unbiased biometric systems. Then, we empirically demonstrate that biometric systems are unfair. Moreover, we highlight that the dataset used to train these systems reproduce the racialisation of subjects, proposing race categories that are archaic and offensive and gender categories that are binary. 

We argue that a critical questioning of the use of fairness in biometrics systems should also be focused on the historical, political and social contexts in which biometrics are deployed. Moving our discourse towards a ‘critical biometric consciousness’~\cite{browne2015dark}, we analyse the case of an asylum appeal where the migrant's credibility was challenged in part due to an inconsistency between his testimony and a biometric trace stored in a database. Given this fact and other inconsistencies, the UK Deputy Upper Tribunal Judge denied his asylum claim, dismissing any other evidence provided by the asylum seeker. This case, we argue, unveils \textit{the elephant in the room} of biometrics: the obvious fact that is being intentionally ignored or left unaddressed, showings how biometrics implemented at the border cannot be fair given that borders' intentional function is to discriminate. While academics, private companies and biometric engineers are centred in building more accurate and fairer biometrics, little attention is paid about how biometrics are jeopardising fundamental rights. Importantly, whilst the European Commission has proposed the first legal framework for artificial intelligence and biometrics, the Artificial Intelligence Act (AI Act hereinafter), to  protect European citizens' digital and fundamental rights~\cite{EC2021AIAAct}, some biometric systems used at the border will be explicitly exempt from such regulation. Therefore, migrants' digital and fundamental rights will not obtain the same lawful protection. Since fairness in biometrics has the risk of becoming more prominent in the incoming years, we urge for a critical and radical examination of this field. We suggest that we must also engage and acknowledge the politics and situated current debates in biometrics within a broaden historical context of struggles against discrimination at the border, moving beyond the technological ethical dilemmas about ethics and biases. To the best of our knowledge, this is the first academic work that investigates fairness in biometrics from a critical perspective, and pushes the argument further showing how debates about the core function of borders and the use of biometrics to criminalise migration are undermined by the focus on ethics and more equitable biometric systems.

\section{Fairness in biometrics: An emergent challenge?}

In 2018, Buolamwini and Gebru~\cite{buolamwini2018gender} published \textit{Gender Shades}, an academic work that assessed bias in gender classification algorithms through facial recognition. They analysed several commercial gender classification models and found significant disparities based on individuals' characteristics: white skin had better results than dark skin whilst males obtained better results than females. This intersectional benchmark opened up a new avenue of critical discourse towards biases in racialised technologies. It influenced public and academic debates towards the use of facial recognition~\cite{kantayya2020coded, benjamin2019race, o2020facial}, creating awareness about the risks of algorithms that encode and propagate historical, political and social biases. Consequently, it  also disrupted the field of biometrics. Since the publication of \textit{Gender Shades}~\cite{buolamwini2018gender}, fairness has emerged as a major challenge within biometrics~\cite{drozdowski2020demographic}.

The main goal of fairness in the field of biometrics is to estimate and mitigate bias in systems that determine the identity or other characteristics like the gender of individuals~\cite{ross2019some}. Researchers have analysed demographic biases in facial recognition, fingerprints, palmprints, iris and even in finger veins. Face recognition has been notably the most analysed system in the last decade, where gender and race are the features analysed. In general, males and white skins obtain higher biometric performance~\cite{lohr2018facial, acien2018measuring, serna2021insidebias, de2020fairness}. The annual reports published by National Institute of Standards and Technology (NIST)~\cite{grother2019face} also found error disparities based on gender and race on more than 189 commercial facial recognition systems that are used for border control. Similarly, other biometrics systems such as iris recognition find noteworthy differences between females and males, with the former having higher error rates~\cite{fang2021demographic}. Biases also persists in fingerprints technologies, but rather than gender and race, age is the demographic feature most affected. The conclusion achieved is that fingerprint verification systems get higher error rates on children~\cite{marasco2019biases, preciozzi2020fingerprint}. Indeed, most of the studies focus on analysing race and gender rather than age. In this case, the performance of some biometric systems such as fingerprints, iris of finger vein might be more affected by age. More interestingly, researchers have suggested that ‘statistically significant bias’ on age and gender ‘have not been detected on five finger vein recognition algorithms tested on four datasets’~\cite{drozdowski2020demographic}.

In spite of the numerous biometric articles in fairness that have been recently published, there is a notable disengagement from relevant and previous works in fairness. First, these works analyse algorithmic bias without considering the more than 20 definitions in fairness in machine learning that have been proposed in the last decade~\cite{verma2018fairness, dunkelau2019fairness, barocas-hardt-narayanan}. For instance, in~\cite{drozdowski2020demographic} demographic bias is calculated by analysing differences of score distributions by groups (males vs. females or children vs. adults). However, differences on error rates which is an standard approach to evaluate fairness are not considered. Second, despite of the popularity of \textit{Gender Shades}~\cite{buolamwini2018gender} and its proposed benchmark, intersectionality is not considered in any study analysed. In general, bias is analysed taking only a single demographic feature into account (gender, race or age). Third, some studies also suggest a lack of bias of biometric systems such as in ~\cite{drozdowski2020demographic}. However, different fairness works have demonstrated the impossibility of fairness, proving mathematically and empirically that fairness definitions are mutually exclusive~\cite{garg2020fairness, chouldechova2017fair,  zhao2019inherent, kleinberg2018inherent}. As a result, it has been recently proved that despite the efforts on debiasing biometric system bias still persists~\cite{grother2019face}.

Part of the literature of fairness in biometrics that we have examined ignores that these systems are nowadays impacting on fundamental rights~\cite{castelvecchi2020beating}. For instance, the authors of \textit{Gender Shades} clearly exposed that ‘all evaluated companies provide a ``gender classification'' feature that uses the binary sex labels of females and male. This reductionist view of gender does not adequately capture the complexities of gender or address transgender identities’~\cite{buolamwini2018gender}. In fact, the mere idea of gender classification algorithms infringes fundamental rights by designing and targeting trans people which are more likely to be ‘miss-classified’.
Examining the performance of classification models with respect to gender will likely improve these models. Yet, we ought to consider if it is even a desirable goal. What is the benefit of gender classification algorithms bring to our societies? In this paper we focus the critique towards the use of biometrics for migration control which impacts on asylum seekers' rights. Since 2000, the EU has implemented biometrics systems to identify travellers for migration control~\cite{scheel2013autonomy, Metcalfe_Dencik_2019, amnesty2016}. Judges, officials in migration administration and border guards among others are currently making use of biometric technology in asylum cases or visa applications~\cite{glouftsios2021inquiry, jones2019picum} to make ‘unequivocal’ decisions. The output of the biometric systems are considered more reliable than migrants or asylum seekers, which are perceived as a deceptive subjects. Thus, the algorithmic performance and error rates are never scrutinised.  

Fairness and the study of bias has become the \textit{elephant in the room} of biometrics. While the effort is focused on solving the challenge of addressing demographic biases and  design ‘fair’ biometric systems, little attention is paid on the context of how it is used. These systems are nowadays reproducing political injustices which are linked to a colonial legacy which is ignored in contemporary biometrics~\cite{browne2015dark}. As Maguire argued~\cite{maguire2009birth}, fingerprints were used by a Brisith officer at the Indian Civil Service to avoid fraud on colonial subjects. At the same time, Galton investigated ‘the heritable characteristics of race’ on racialised individuals in British prisons. In fact, biometrics ‘offered 19th-century innovators more than the prospect of identifying criminals: early biometrics promised a utopia of bio-governmentality in which individual identity verification was at the heart of population control’~\cite{maguire2009birth}. In the 21th-century, this approach to biometrics has evolved with the use of algorithms that automatically calculate matches. Yet its colonial legacy remains still visible. At the border, these systems are used to identify racialised subjects. There, fairness has emerged to make the identification of asylum seekers ‘fairer’.

\section{A theoretical approach towards fairness in biometrics}
Fairness is a contested concept that have been historically discussed by social scientists, legal experts and philosophers. Yet there is no consensus on what this concept means. This idea have been translated into mathematical definitions which quantify the fairness of algorithmic-based models~\cite{hutchinson2019unfairness}, resulting in different formulations~\cite{verma2018fairness, dunkelau2019fairness, barocas-hardt-narayanan}. In recent years, the research field in biometrics have also evaluated the bias and fairness of these systems as well as developing novel methods for bias mitigation. Most of the studies that we have analysed do not take into account the most relevant state-of-the-art definitions of fairness in machine learning, and some of them argued that their experimental evaluation suggests lack of bias in score distributions~\cite{drozdowski2021demographic}. Moreover, recent studies of fairness in machine learning have proved to be generally impossible to satisfy several fairness criteria simultaneously~\cite{garg2020fairness, chouldechova2017fair,  zhao2019inherent, kleinberg2018inherent}. 

In this section, we mathematically demonstrate the impossibility of bias-free biometric systems, showing that several fairness criteria can only be satisfied under very limited conditions. We introduce a formulation for biometric verification systems and translate fairness criteria into this formulation to finally demonstrate the incompatibility among them.

\subsection{Formulation of a biometric verification system}

In the discussion herein we focus on biometric verification systems. These systems confirms (or rejects) whether a biometric sample belongs to an specific individual based on similarity to their learned representation. Formally, we use the following notation:

\begin{itemize}
    \item $x^{(i)}$: learned representation of the biometric $i$th-sample within the dataset $\mathcal{D}$,
    \item $y^{(i,j)}$: element $(i,j)$ of the binary outcome vector $y$,
    \item $\theta$: set of parameters of the biometric system,
    \item $\tilde{y}^{(i,j)} = f ((x^{(i)}, x^{(j)}) | \theta)$:  element $(i,j)$ of the binary prediction vector $\tilde{y}$,
    \item $\tau$: decision threshold,
    \item $s^{(i,j)}$: score of similarity between samples $i$ and $j$.
\end{itemize}

These systems employ deep learning algorithms which transform images into numerical representations. They are driven by an optimisation function to minimise errors between two samples ($x^{(i)}$ and $x^{(j)}$) and the outcome ($y^{(i,j)}$) which is whether they belong to the same individual or not:

\begin{equation}
\label{eq:optimisation}
\theta^\ast = \argmin_{\theta} \sum_{x^{(i)},x^{(j)} \in \mathcal{D}} \mathcal{L} \Big( f \left( (x^{(i)}, x^{(j)}) \big| \theta \right),  y^{(i,j)} \Big)
\end{equation}

The biometric model compares two learned representations samples ($x^{(i)}$ and $x^{(j)}$) and obtains a similarity score ($s^{(i,j)}$). This score is then translated into a binary prediction ($\tilde{y}^{(i,j)}$) given a threshold $\tau$:

\begin{equation}
\label{eq:decision}
\tilde{y}^{(i,j)} =
    \begin{cases}
     0  & \text{if $s \leq \tau$,} \\
     1 & \text{if $s > \tau$} 
    \end{cases}   
\end{equation}

The model learns the best representation of parameters ($\theta$) that achieves the minimum number of ‘miss-identifications’. This means that the  binary prediction ($\tilde{y}^{(i,j)}$) should be as similar as the binary outcome ($\tilde{y}^{(i,j)}$). In the case of biometric verification systems, two pairs of biometric samples are labelled as \textit{genuine} ($y^{(i,j)}=1$) if they correspond to the same individual and as \textit{impostor} ($y^{(i,j)}=0$) otherwise~\footnote{Note the language used in biometrics: a miss-match is defined as an \textit{impostor} which denotes that identity's deception is taken for granted. In the machine learning literature, there is an absence of critical, transdisciplinary and genealogical examination of this term an other concepts~\cite{birhane2021impossibility}.}. Therefore, a \textit{true genuine} ($TG = Pr(\tilde{y}^{(i,j)} = 1 | y^{(i,j)} = 1)$) refers to those samples that correspond to the same individual and are correctly matched; the \textit{true impostor} ($TN = Pr(\tilde{y}^{(i,j)} = 0 | y^{(i,j)} = 0)$) refers to samples that correspond to different individuals  and they are rejected; \textit{false genuine} ($FG = Pr(\tilde{y}^{(i,j)} = 1 | y^{(i,j)} = 0)$) refers to different individuals whose samples are matched; and \textit{false impostor} ($FI = Pr(\tilde{y}^{(i,j)} = 0 | y^{(i,j)} = 1)$) refers to samples of the same individual that are miss-matched\footnote{In the biometric literature, false genuine and impostor rates are also known as false match (FMR) and non-match (FNMR) rates respectively.}. Thus, incorrect association of two subjects or failed association of one subject are the errors that biometric systems commit. Formally, the true genuine rate ($TGR$), true impostor rate ($TIR$), false genuine rate ($FGR$), and false impostor rate ($FIR$) are formulated as:

\pgfmathdeclarefunction{gauss}{2}{\pgfmathparse{1/(#2*sqrt(2*pi))*exp(-((x-#1)^2)/(2*#2^2))}}
\begin{center}
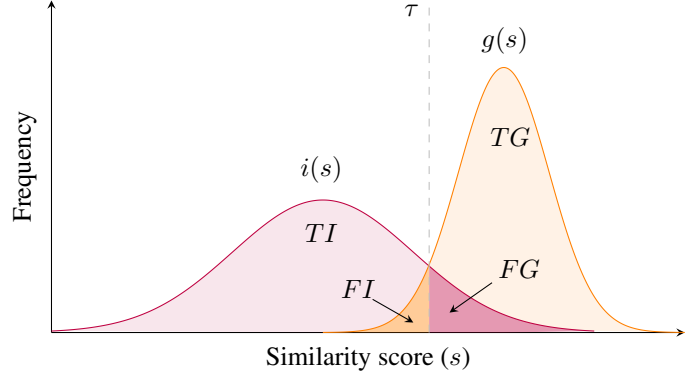
\begin{figure}
\centering
\begin{tikzpicture}
\begin{axis}[
no markers, domain=0:20, samples=100,
xlabel=Similarity score ($s$), ylabel=Frequency,
height=6cm, width=10cm,
axis lines = left,
xtick = \empty, 
ytick = \empty,
clip = false,
]

\centering
\addplot[domain = 0:10, domain=-3:3, samples = 400, color = purple, name path = genuine]{gauss(0,1)};
\addplot[domain = 0:10, domain=0:4, samples = 400, color = orange, name path = impostor]{gauss(2,0.5)};
\addplot [domain = 0:10, domain = 0:2, draw = none, name path = xaxis]{0};

\centering
\addplot[color = lightgray, dashed, name path = tau] coordinates {(1.175, 0) (1.175, 1)};

\node[right] at (380, 97) {$\tau$};
\node[right] at (270, 30) {$TI$};
\node[right] at (265, 49) {$i(s)$};
\node[right] at (475, 59) {$TG$};
\node[right] at (465, 88) {$g(s)$};
\node[right] at (310, 13) {$FI$};
\node[right] at (484, 19) {$FG$};

\draw [-stealth](355, 10) -- (400, 3);
\draw [-stealth](487, 15) -- (440, 6);

\addplot [orange, opacity = 0.4] fill between [of = impostor and xaxis, soft clip={domain=-3:1.175}];
\addplot [purple, opacity = 0.4] fill between [of = genuine and xaxis, soft clip={domain=1.175:3}];
\addplot [purple, opacity = 0.1, dashed] fill between [of = impostor and genuine, soft clip={domain=-3:1.175}];
\addplot [orange, opacity = 0.1, dashed] fill between [of = genuine and impostor, soft clip={domain=1.175:5}];

\end{axis}
\end{tikzpicture}
\caption{Genuine ($g(s)$) and impostor ($i(s)$) score distributions for a biometric verification system. The figure shows the relation between score distributions and the confusion matrix elements (true genuine ($TG$), true impostor ($TI$), false genuine ($FG$), and false impostor ($FI$)).}
\label{fig:distributions}
\end{figure}
\end{center}

In contrast to machine learning, biometric systems are evaluated setting different decision thresholds ($\tau$) which clearly affects on the distribution of errors (see Figure~\ref{fig:distributions}). Equal Error Rate ($EER$) is the value where the false genuine and impostor rates curves intersects ($FGR = FIR$). False genuine rate at 0.001 ($FGR_{1000}$) is the value of false impostor rate when false genuine rate is 0.001 ($FGR = 0.001$), and vice versa. Systems are also evaluated when one of these rates is 0 ($ZFGR$ when $FIR=0$ or $ZFIR$ when $FGR=0$). Thus, the decision threshold is set targeting different values of error.

\subsection{Formulation of three fairness definitions for biometrics}

A fairness measure is a mathematical function that quantifies and assesses biased systems and algorithmic discrimination. These measures aimed at evaluating model performance across demographics groups ($\mathcal{C} = \{C_1, C_2, \ldots, C_n\}$) and ensure that there is no disparate impact among them. In this setting, typically an individual feature such as gender, class or ethnicity is proposed to define the ‘advantaged’ and ‘disadvantaged’ group. 

We detect that the current literature of fairness in biometrics disengages from the state-of-the-art of fairness in machine learning. In most of the works analysed~\cite{drozdowski2021demographic, drozdowski2020demographic, terhorst2021comprehensive, fang2021demographic}, authors assessed biases in biometric systems by proposing their own fairness definitions without considering previous works on algorithmic discrimination. Therefore, we propose three well-known definitions of fairness translated into biometrics.

\subsubsection{\textbf{Equalised odds}  (also \textit{disparate mistreatment} or \textit{error rate balance})~\cite{hardt2016equality}} A biometric system satisfies \textit{equalised odds} if TGR and FGR are similar across demographics groups:

\begin{equation}
\label{eq:tpr_eqod}
| TGR_{C_a} - TGR_{C_b} | < \epsilon
\end{equation}
and, 

\begin{equation}
\label{eq:fpr_eqod}
| FGR_{C_a} - FGR_{C_b} | < \epsilon, \forall~C_a, C_b \subset \mathcal{C}
\end{equation}

\subsubsection{\textbf{Statistical parity} (also \textit{group fairness} or \textit{demographic parity})~\cite{dwork2012fairness}}
A biometric system satisfies \textit{statistical parity} if the probability of predicted genuine is similar across demographics groups. This definition is based on the predicted outcome. Mathematically, this definition is expressed as follows:

\begin{equation}
\label{eq:stat_par}
|Pr(\tilde{y} = 1 |  C_a ) -  Pr(\tilde{y} = 1 |  C_b )| < \epsilon, \forall~C_a, C_b \subset \mathcal{C}
\end{equation}

\subsubsection{\textbf{Predictive parity} (also \textit{ outcome test})~\cite{chouldechova2017fair}}
A biometric system satisfies \textit{predictive parity} if the probability of being predicted genuine of actual genuine is similar across demographic groups. More formally:

\begin{equation}
\label{eq:pred_par}
|Pr(y = 1 |  \tilde{y} = 1, C_a ) -  Pr(y = 1 | \tilde{y} = 1, C_b )| < \epsilon, \forall~C_a, C_b \subset \mathcal{C}
\end{equation}

\subsection{The impossibility of unbiased biometric systems}
In this section, we provide a theoretical framework to demonstrate that the previous fairness criteria are mutually exclusive. Building on previous proofs~\cite{garg2020fairness, chouldechova2017fair,  zhao2019inherent, kleinberg2018inherent}, we mathematically show that only under very unrealistic conditions (equal ratios among demographics groups, trivial or perfect biometric system~\footnote{The trivial biometric system is the system than always outputs one class (genuine or impostor). The perfect biometric system is an utopic system that achieves zero classification error, i.e., $FGR=0$ and $FIR=0$.}), these three definitions can be simultaneously satisfied. Thus, we prove the impossibility of any unbiased biometric system.

To simplify the notation, we assume that $\mathcal{C} = \{C_1, C_2\}$.

\begin{prop}
\label{prop:pred_par_2}
Given a biometric system which is non-trivial with unequal ratios among groups that satisfies equalised odds and statistical parity, then predictive parity cannot hold.
\end{prop}
\begin{proof}

Given Bayes' theorem, we obtain that:

\begin{equation}
\label{eq:bay_th_fair}
Pr(y=1 | \tilde{y}=1, \mathcal{C}) = \frac{Pr(\tilde{y}=1 | y=1, \mathcal{C}) Pr(y=1 | \mathcal{C})}{Pr(\tilde{y}=1 | \mathcal{C})}
\end{equation}

If predictive parity is satisfied, then:

$$|Pr(y = 1 |  \tilde{y} = 1, C_1 ) -  Pr(y = 1 | \tilde{y} = 1, C_2 )| < \epsilon \overset{\text{Equation}~ \ref{eq:bay_th_fair}}{\implies} $$

$$ \Big| \frac{Pr(\tilde{y} = 1 | y = 1,  C_1) Pr(y=1 | C_1) }{Pr(\tilde{y} = 1 |C_1)} - \frac{Pr(\tilde{y} = 1 | y = 1,  C_2) Pr(y=1 | C_2) }{Pr(\tilde{y} = 1 | C_2)} \Big |   < \epsilon $$

Assuming that equalised odds on the TGR ( $| TGR_{C_1} - TGR_{C_2} | < \epsilon)$ and statistical parity ($|Pr(\tilde{y} = 1 |  C_a ) -  Pr(\tilde{y} = 1 |  C_b )| < \epsilon$) are satisfied in the previous expression:

$$
\Big| \frac{Pr(\tilde{y} = 1 | y = 1) [Pr(y=1 | C_1)-Pr(y=1 | C_2)]}{Pr(\tilde{y} = 1)} \Big |   < \epsilon
$$

Given that $Pr(\tilde{y} | y ), Pr(y| \mathcal{C}), Pr(\tilde{y}) \in [0, 1]$, predictive parity is only satisfied when:

$$ | Pr(\tilde{y} = 1 | y = 1) [Pr(y=1 | C_1)-Pr(y=1 | C_2)] |  < \epsilon$$

which implies:

$$ Pr(\tilde{y} = 1 | y = 1) < \epsilon  $$

or

$$ | Pr(y=1 | C_1)-Pr(y=1 | C_2) | < \epsilon . $$

On one hand, if $ Pr(\tilde{y} = 1 | y = 1) < \epsilon $, we obtain that Proposition~\ref{prop:pred_par_2} is only satisfied when $TGR = 0$ which  means that the system fails to correctly classified any genuine instance or the system is a trivial classifier, e.g. there are only impostor instances. On the other hand,  $ | Pr(y=1 | C_1)-Pr(y=1 | C_2) < \epsilon | \implies  Pr(y=1 | C_1) \sim  Pr(y=1 | C_2)$ which implies that Proposition~\ref{prop:pred_par_2} is satisfied only under equal ratios. 
\end{proof}

\section{A biometric experiment: Why this system is and will be always biased?}

We examine four biometric systems (finger veins) to empirically demonstrate the impossibility of the impossibility of fairness (see Propostion~\ref{prop:pred_par_2}). These biometric algorithms identify subjects through vascular patterns on the human body, i.e. finger, palm or human eye veins~\cite{uhl2020handbook}. These four systems are proposed in~\cite{drozdowski2021demographic} to assess demographic bias  by differences on score distribution statistics (mean and standard deviation) of genuine and impostor attempts are evaluated. The conclusion achieve is that statistically significant biases in score distributions do not exist and the authors proposed to evaluate this framework in the future with more individuals, given that the number of subjects in each of the databases is very small. Rather than reproduce their experiments with larger databases, we empirically demonstrate that this framework is biased from three different perspectives: (1) ratios, (2) fairness criteria, and (3) intersectionality analysis. 

\subsection{Ratios}

\begin{figure}[!htbp]
  \centering
  \begin{minipage}[b]{0.47\textwidth}
    \includegraphics[width=\textwidth]{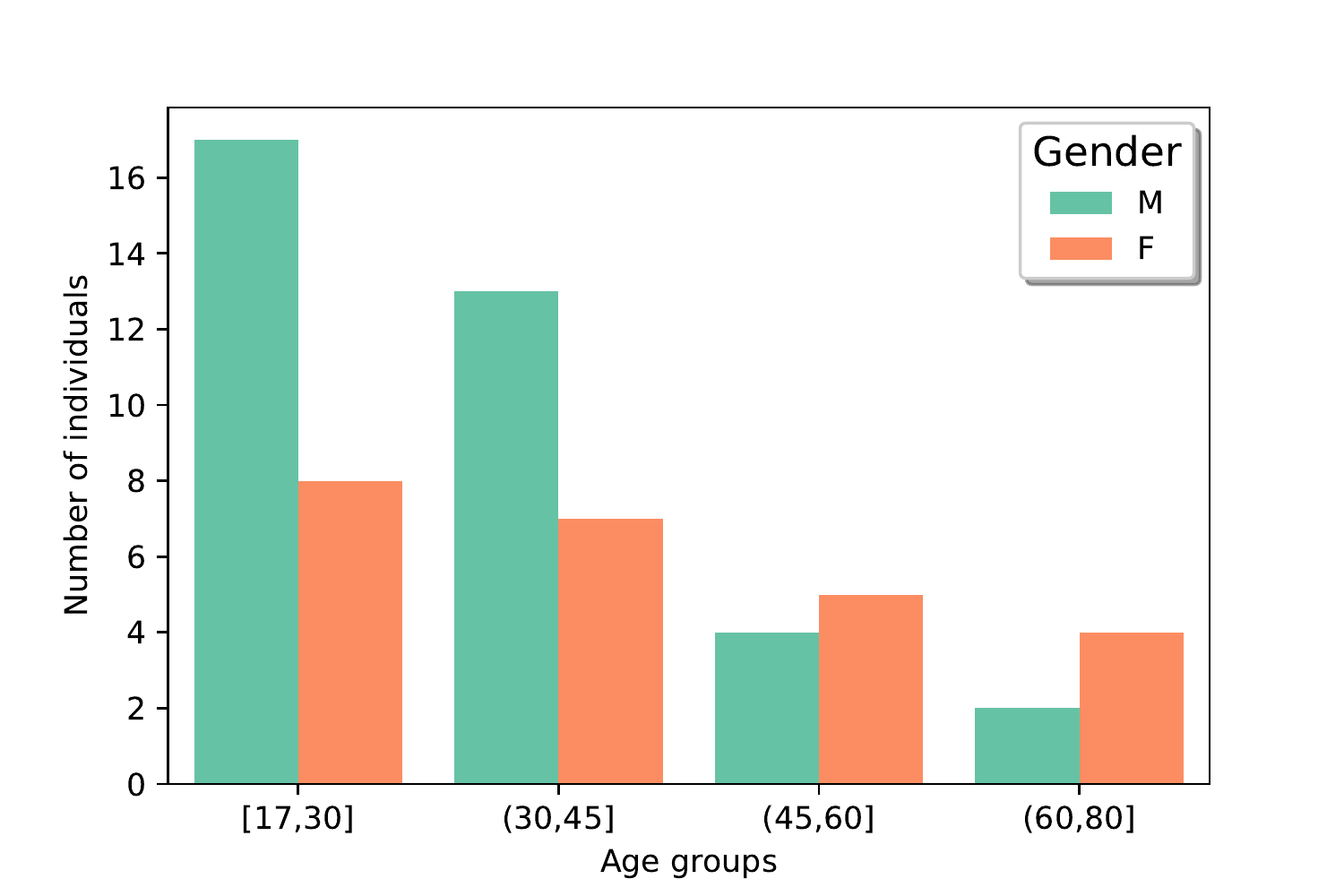}
  \end{minipage}
  \hfill
  \begin{minipage}[b]{0.47\textwidth}
    \includegraphics[width=\textwidth]{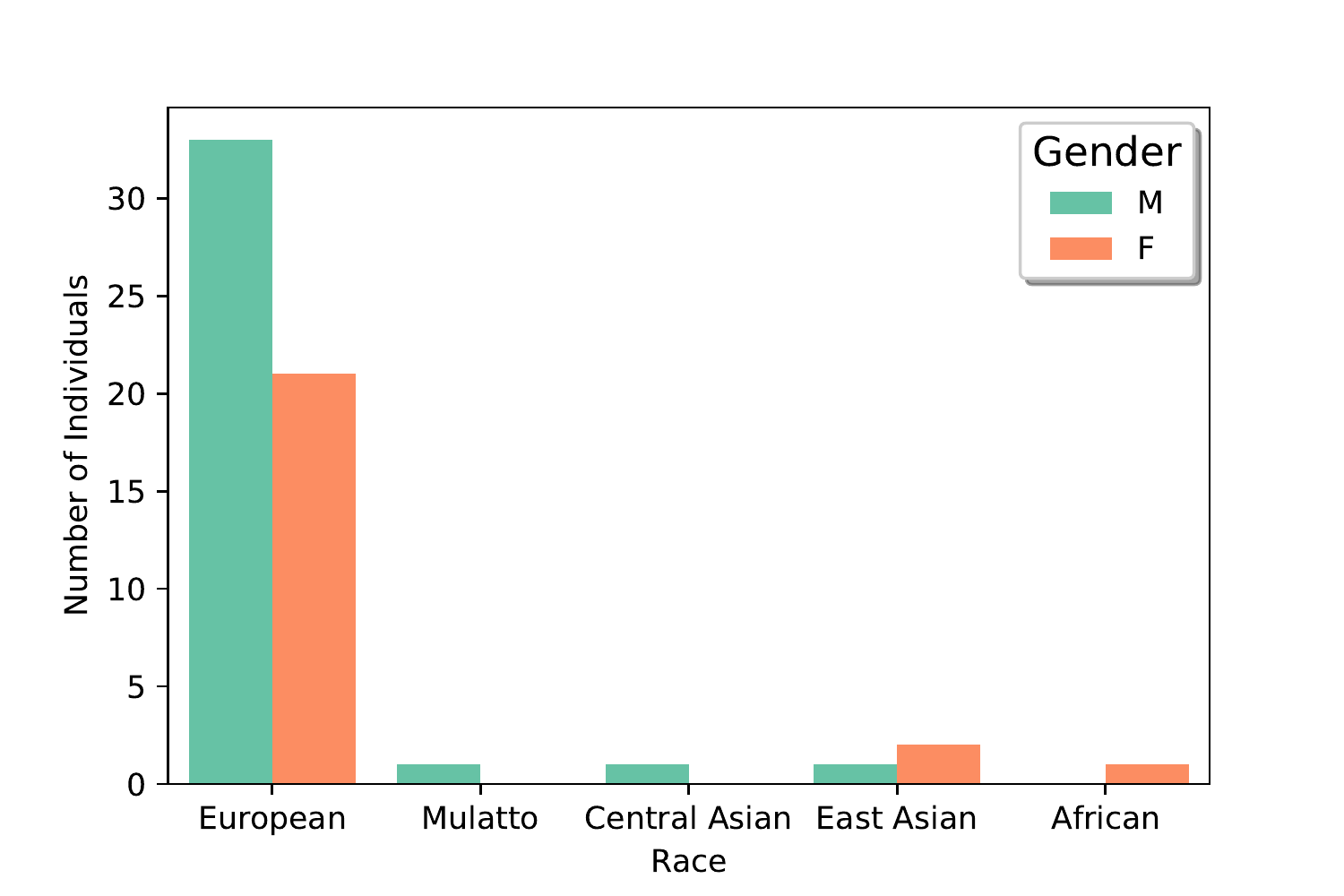}
  \end{minipage}
   \caption{Intersectional ratios on age, ethnicity and gender of individuals in PLUSVein-FV3~\cite{kauba2018focussing}. There are large disparities among groups: more males (M) than females (F), young than old adults, and majority of Europeans. The proposed ethnicity labels lack of diversity, considering one of them (‘Mulatto’) rather archaic and offensive.}
   \label{fig:gender}
\end{figure}

Broken links and APIs hampered the access to three of four publicly available datasets~\cite{lu2013available, ton2013high, vanoni2014cross}. Through an online petition\footnote{See: \url{https://wavelab.at/sources/PLUSVein-FV3} (accessed 06 December 2021)}, we had access to PLUSVein-FV3\footnote{The dataset provided is completely anonymised. There is no possibility that the direct linking of this information to an individual could lead to their identification. Our sole research intention in processing this data is to demonstrate the impossibility of fairness in biometric systems, which is in the public interest.}~\cite{kauba2018focussing}. This database contains 1440 finger vein images of 60 individuals from different hands and fingers. Figure~\ref{fig:gender} shows the number of individuals based on age, race\footnote{In the original work, the authors used the ethnicity concept. We propose to rather use the race category which is related to the colonial and legal construction of human beings categories.} and gender\footnote{Note that in the original work, the authors used sex instead of gender. The authors argued in~\cite{drozdowski2020demographic} that: ‘[T]erms ``gender'' and ``sex'' are often used in a binary and conflated manner’. Moreover, they proposed to use ISO/IEC's definitions that distinguish biological sex from cultural gender. Building on Butler's claim that there are no distinctions between sex and gender~\cite{butler1999gender}, we propose to use the concept gender.}. We observe that the mean of age within this dataset is 37.9 years, whilst  $Q_3$ is 46.5 years, which implies that the database is not representative for elders. Europeans are the most represented race in PLUS-VeinFV3: 90\% European, 5\% East Asian, 1.6\% Central Asian, 1.6\% ‘Mulatto’, 1.6\% African. The proportion of non-European individuals is significantly low (see Figure~\ref{fig:gender}). Remarkably, ‘Mulatto’ is a label proposed for this demographic which is not linked to any continent. Whilst European, East Asian and Central Asian correspond to race labels related with geographical expressions, ‘Mulatto’ was a label used during the Spanish colonial period to mark the slave status of children born to Spaniards and African women slaves. This category exposes the colonial and racial legacy of biometrics through a conceptualisation of racialised and colonial bodies. ‘Mulatto’ is a clear expression of the making of bodies through their qualities of ‘colour’ and colonial slavery~\cite{browne2015dark, gilroy2000against}. Analysing the gender feature, we observe that the dataset is more balanced: 60\% males and 40\% females. Yet this feature does not consider other gender expressions rather than binary ones.

Bias on demographic groups in PLUSVein-FV3 becomes evident in Figure~\ref{fig:gender}. Any biometric system trained in this dataset will perform better on individuals whose demographic characteristics are widely presented in the database, i.e. young male Europeans. The performance of the biometric system will be poor on individuals who are under-represented: non-Europeans, elderly and females. Consequently, given that the majority of samples are taken from European males, rates of genuine and impostor outcomes will be significantly unequal across different demographic groups. Thus, as previously demonstrated (Proposition~\ref{prop:pred_par_2}) equalised odds, statistical parity and predictive parity cannot hold together.

\subsection{Fairness criteria}

The experiments are run using four finger vein recognition systems. These systems are designed using different types of vein recognition schemes: LBP~\cite{feng2016finger}, MC~\cite{miura2007extraction}, PC~\cite{choi2009finger}, and SIFT~\cite{xie2019finger, kauba2014pre}. Table~\ref{tab:error_metrics} shows the results on the PLUSVein-FV3. In general, the error rates of these systems are very low (see MC, PC and SIFT). However, we observe that LBP has the worst performance, obtaining 0.89 and 0.78 of $FIR$ at $FGR_{1000}$ and $FGR_{100}$ respectively. In this case, if the rate of false match is very low ($\downarrow FGR$), the false non-match rate is extremely high ($\uparrow FIR$). Nevertheless, the aim of this section is to demonstrate that biometric systems are and will be always biased. To do so, we calculate three fairness criteria (equalised odds, statistical parity and predictive parity) on three different demographics (age, gender, and ethnicity). Demographics groups are categorised as: young ($\leq45$) and old ($>45$), male and female, and European and non-European.

\begin{table}[]
\centering
\begin{tabular}{lllllll}
 &  & \multicolumn{4}{c}{\textbf{Performance metric}} \\ \cline{3-7} 
\textbf{} &  & $EER$ & $FGR_ {1000}$ & $FGR_{100}$ & $FGR_{10}$  & $ZFIR$ \\ \cline{2-7} 
\multirow[c]{4}{*}{\rotatebox[origin=tr]{90}{\textbf{Method}}} & LBP & 0.16 & 0.89 & 0.78 & 0.20 & 0.99\\
 & MC & 0.02 & 0.03 & 0.02 & 0.01 & 0.99\\
 & PC & 0.02 & 0.05 & 0.03 & 0.02 & 0.98\\
 & SIFT & 0.02 & 0.04 & 0.02 & 0.01 & 1.00 \\ \cline{2-7} 
\end{tabular}
\caption{Error rates of biometric methods. LBP is the method with the poorest performance. MC, PC, and SIFT obtain low and similar error rates. \label{tab:error_metrics}}
\end{table}

As de Freitas Pereira and Marcel observed~\cite{de2020fairness}, several works of fairness in biometrics set a single $\tau$ for every demographic group. However, they argue that this is a ‘serious flaw’ and ‘can give a false impression that a biometric verification system is fair’~\citep[p.~3]{de2020fairness}. They conclude that ‘[f]air biometric recognition systems are fair if a decision threshold $\tau$ is ``fair'' for all demographic groups with respect to $FGR$($\tau$) and $FIR$($\tau$)’~\citep[p.~10]{de2020fairness}. Following this suggestion, fairness metrics are calculated after setting the same decision thresholds for the three demographic groups: $FGR_{1000}$  and near $ZFIR$.

\begin{table}[]
\begin{tabular}{llllllllll}
 & \multicolumn{3}{c}{\textbf{equalised odds}} & \multicolumn{3}{c}{\textbf{statistical parity}} & \multicolumn{3}{c}{\textbf{predictive parity}} \\ \cline{2-10} 
 &  \multicolumn{1}{c}{age} & \multicolumn{1}{c}{gender} & \multicolumn{1}{c|}{race} & \multicolumn{1}{c}{age} & \multicolumn{1}{c}{gender} & \multicolumn{1}{c|}{race} & \multicolumn{1}{c}{age} & \multicolumn{1}{c}{gender} & \multicolumn{1}{c}{race} \\ \hline
LBP & {[}4.26\%, \textcolor{purple}{40.19\%} {]} & {[}\textcolor{purple}{29.02\%} , \textcolor{purple}{11.36\%}  {]} & \multicolumn{1}{l|}{{[}\textcolor{purple}{44.29\%} , \textcolor{purple}{19.87\%} {]}} & \textcolor{purple}{32.16\%} & \textcolor{purple}{47.91\%}  & \multicolumn{1}{l|}{\textcolor{purple}{45.17\%} } & \textcolor{purple}{11.21\%} & 4.98\% & 3.03\% \\
MC & {[}\textcolor{purple}{7.91\%}, \textcolor{purple}{86.06\%} {]} & {[}4.80\%, \textcolor{purple}{129.57\%} {]} & \multicolumn{1}{l|}{{[}3.60\%, 0.07\%{]}} & \textcolor{purple}{42.03\%} & \textcolor{purple}{16.05\%}& \multicolumn{1}{l|}{\textcolor{purple}{7.24\%}} & 0.46\% & 1.27\% & 0.14\% \\
PC & {[}\textcolor{purple}{9.47\%}, \textcolor{purple}{76.76\%} {]} & {[}3.58\%, \textcolor{purple}{421.18\%} {]} & \multicolumn{1}{l|}{{[}3.96\%, \textcolor{purple}{198.80\%}{]}} & \textcolor{purple}{39.53\%} & \textcolor{purple}{19.12\%} & \multicolumn{1}{l|}{\textcolor{purple}{10.73\%}} & 0.40\% & 2.59\% & 2.68\% \\
SIFT & {[}\textcolor{purple}{6.33\%}, \textcolor{purple}{7.90\%}  {]} & {[}\textcolor{purple}{5.32\%} , \textcolor{purple}{43.35\%}  {]} & \multicolumn{1}{l|}{{[}\textcolor{purple}{5.14\%} , \textcolor{purple}{49.51\%} {]}} & \textcolor{purple}{42.74\%}  & \textcolor{purple}{12.68\%}  & \multicolumn{1}{l|}{\textcolor{purple}{7.91\%} } & 0.74\% & 1.11\% & 0.99\% \\ \hline
\end{tabular}
\caption{Biometric recognition performance as measured by fairness criteria differences among three demographic groups at $FGR_{1000}$. All systems are consistently unfair, showing significant equalised odds and statistical parity differences for age, gender and race.~\label{tab:fairness_metrics_fmr100}}
\end{table}

\begin{table}[]
\begin{tabular}{llllllllll}
     & \multicolumn{3}{c}{\textbf{equalised odds}}                                                        & \multicolumn{3}{c}{\textbf{statistical parity}}                                       & \multicolumn{3}{c}{\textbf{predictive parity}}                                       \\ \cline{2-10} 
     & \multicolumn{1}{c}{age} & \multicolumn{1}{c}{gender} & \multicolumn{1}{c|}{race}              & \multicolumn{1}{c}{age} & \multicolumn{1}{c}{gender} & \multicolumn{1}{c|}{race} & \multicolumn{1}{c}{age} & \multicolumn{1}{c}{gender} & \multicolumn{1}{c}{race} \\ \hline
LBP  & {[}0.36\%, 0.35\%{]}  & {[}0.36\%, 0.42\%{]}     & \multicolumn{1}{l|}{{[}1.01\%, 0.96\%{]}} & 0.36\%                  &      0.51\%          & \multicolumn{1}{l|}{1.01\%}    & \textcolor{purple}{53.22\%}                      & \textcolor{purple}{19.19\%}                      & 3.66\%                       \\
MC     & {[}0.08\%, 1.25\%{]}   & {[}0.42\%, 1.92\%{]}      & \multicolumn{1}{l|}{{[}0.32\%, 1.44\%{]}}  & 1.19\%                  & 1.84\%                    & \multicolumn{1}{l|}{1.32\%}    & \textcolor{purple}{51.58\%}                  & \textcolor{purple}{18.67\%}                  & \textcolor{purple}{5.38\%}                        \\
PC   & {[}0.62\%, 3.55\%{]}    & {[}1.97\%, 9.33\%{]}      & \multicolumn{1}{l|}{{[}0.58\%, \textcolor{purple}{7.24\%}{]}}   &3.38\%                 & \textcolor{purple}{8.88\%}                     & \multicolumn{1}{l|}{\textcolor{purple}{6.65\%}}    & \textcolor{purple}{47.46\%}                 & \textcolor{purple}{12.71\%}                   & \textcolor{purple}{11.69\%}                       \\
SIFT  & {[}3.40\%,  3.45\%{]}   & {[}0.36\%, 0.42\%{]}       & \multicolumn{1}{l|}{{[}1.01\%, 0.96\%{]}}   & 2.79\%                & 0.61\%                    & \multicolumn{1}{l|}{1.01\%}    & \textcolor{purple}{52.47\%}                 & \textcolor{purple}{19.19\%}                   & 3.66\%                       \\ \hline
\end{tabular}
\caption{Biometric recognition performance as measured by fairness criteria differences among three demographic groups at $\sim ZFIR$. All systems are consistently unfair, showing significant predictive parity differences for age, gender and ethnicity. These results empirically demonstrate Proposition~\ref{prop:pred_par_2}.~\label{tab:fairness_metrics_zerofnmr}}
\end{table}

The overall recognition performance results show that the four biometric systems are unfair considering three fairness definitions on the three demographic groups assessed\footnote{We consider that the system is unfair if the difference is greater than 5\%.}. These results contradicts the idea that these systems lack of demographic bias. Moreover, the results on Table~\ref{tab:fairness_metrics_zerofnmr} hold our theoretical framework on the impossibility of unbiased biometric systems (Proposition~\ref{prop:pred_par_2}). We observe how the decision threshold value ($\tau$) impacts on the fairness criteria that are satisfied. On one hand, setting $\tau$ at $FGR_{1000}$ both equalised odds and statistical parity are not satisfied, yet predictive parity is achieved (Table~\ref{tab:fairness_metrics_fmr100}). On the other hand, when $\tau$ is set at $\sim ZMR$ then predictive party is held but equalised odds and statistical parity are not achieved (Table~\ref{tab:fairness_metrics_zerofnmr}).

The empirical results at $FGR_{1000}$ clearly hold Chouldechova's observation in~\citep[p.~157]{chouldechova2017fair} about predictive parity, $FGR$, and $FIR$. When a system satisfies predictive parity but ratios differ across demographic groups, the system cannot achieve equal FGR and FIR across those groups. Thus, if FGR are not similar equalised odds cannot be satisfied. Analysing disparities among groups, old adults obtain worsen results than young, which could be a consequence of low proportion of old adults in the dataset. Gender also has significant unfair results. For instance, MC and PC obtains higher $FGR$ on females than on males (421.18\% and 198.8\% respectively). Fair criteria are also worsen on non-Europeans than Europeans. 

On the other hand, setting thresholds near $ZFIR$ ($\sim ZFIR$) implies that predictive parity is not satisfied. These results clearly demonstrate Proposition~\ref{prop:pred_par_2} that states that the three fairness definition cannot hold simultaneously\footnote{Note we cannot set the threshold at $ZFIR$ because this implies dividing by zero when calculating fairness definitions.}. Rather than ethnicity, we observe that age and gender obtain wider differences. In this case, old people and female individuals obtain larger error rates than young and males respectively.



\subsection{Intersectionality}

\begin{figure}[!h]
    \centering
    \begin{subfigure}{0.5\textwidth}
        \includegraphics[width=1\textwidth]{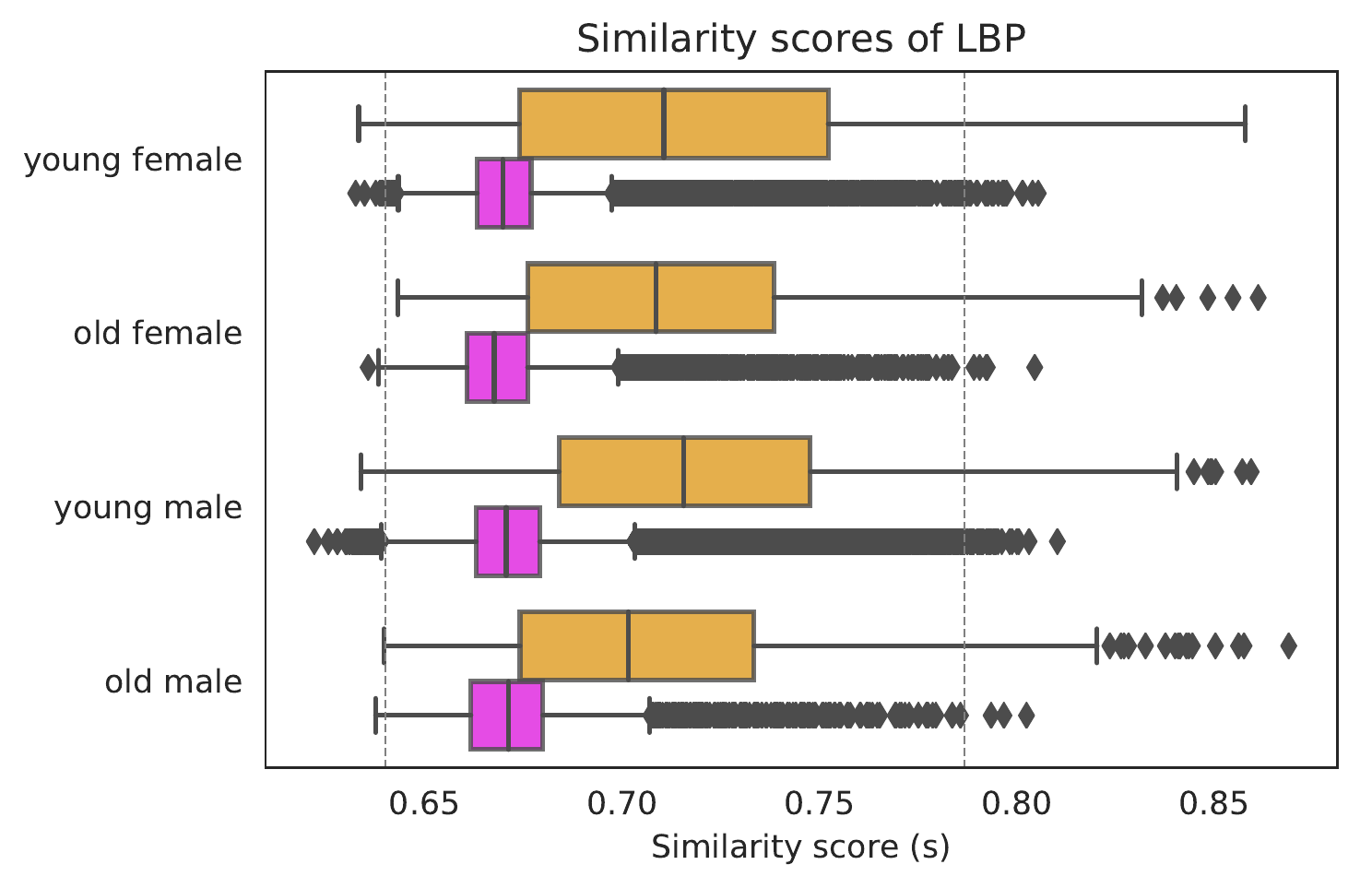}\vspace{-2mm}
        \subcaption{LBP}
    \end{subfigure}\hfill
    \begin{subfigure}{0.5\textwidth}
        \includegraphics[trim=0 0 0 0,width=1\textwidth]{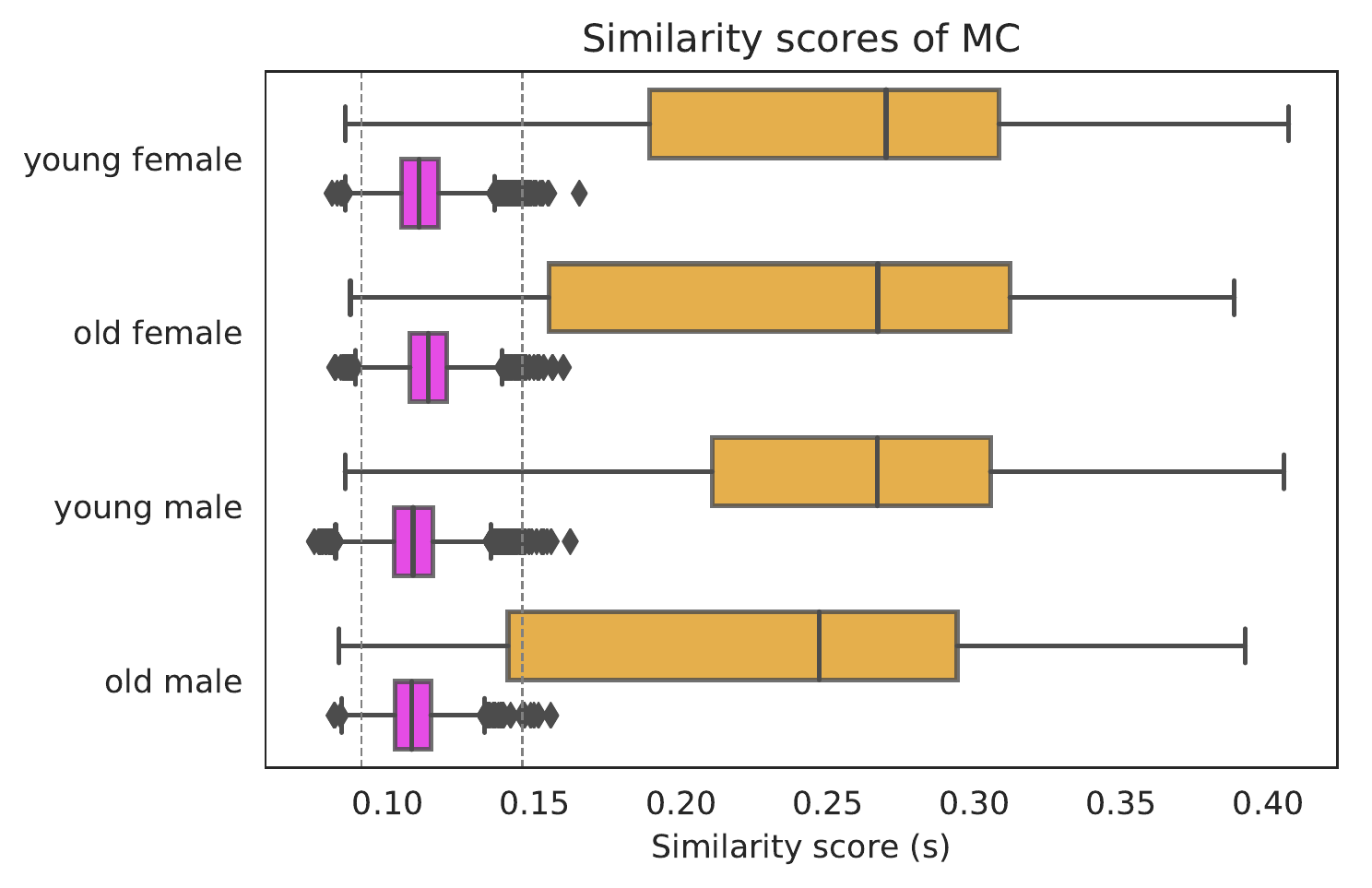}\vspace{-2mm}
        \subcaption{MC}
    \end{subfigure}\\[2mm]
    \begin{subfigure}{0.5\textwidth}
        \includegraphics[trim=0 0 0 0,clip,width=\textwidth]{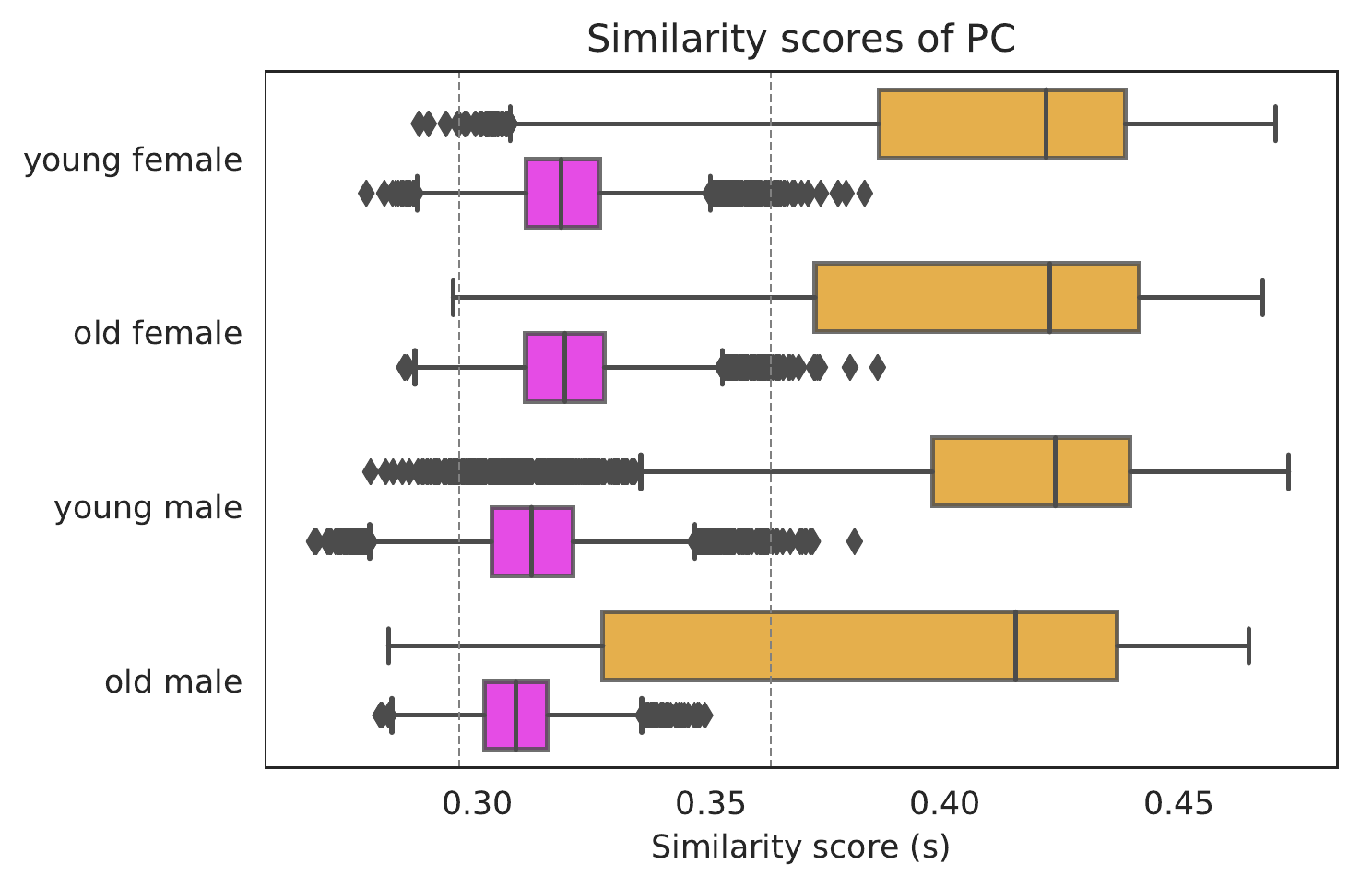}\vspace{-2mm}
        \subcaption{PC}
    \end{subfigure}\hfill
    \begin{subfigure}{0.5\textwidth}
        \includegraphics[trim=0 0 0 0,width=1\textwidth]{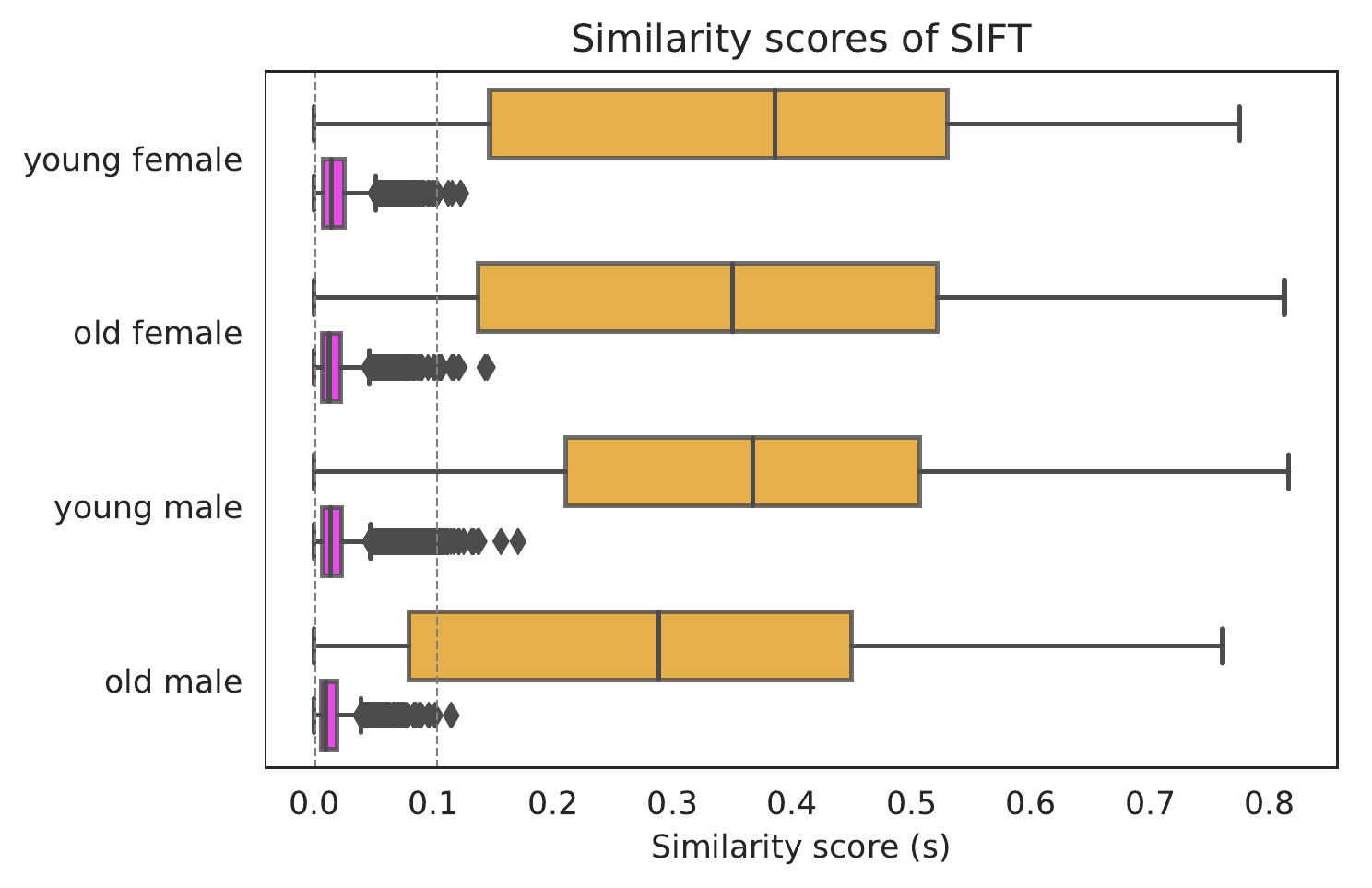}\vspace{-2mm}
        \subcaption{SIFT}
    \end{subfigure}
    \caption{Intersectional fairness disparities in four biometric recognition systems based on age and gender. Decision threshold ($\tau$) is set at $\sim ZFIR$ (first dashed line) and $FGR_{1000}$ (second dashed line). Distribution of genuine (orange) and impostor (purple) scores differ among groups. Classification disparities across intersectional demographic groups are apparent. Overall, young males obtain better performance than other groups.}
    \label{fig:intersectional_boxplots}
\end{figure}

Given the intersectional benchmark proposed in \textit{Gender Shades}~\cite{buolamwini2018gender}, we evaluate the results of the four biometric recognition systems based on intersectional groups. The results show that disparities are more prominent on age and gender, rather than on ethnicity. Unlike facial recognition, finger veins systems do not taken into account skin tones and their performance is more affected by other attributes such as size of fingers. Moreover, wide differences in ratios among Europeans and non-Europeans (Figure~\ref{fig:gender}) impacts also on these results.

Figure~\ref{fig:intersectional_boxplots} shows the distribution of genuine and impostor scores across four groups (young females, old females, young males and old males) at two decision thresholds. On the four biometric systems, young male scores obtain better results than the other groups. The distribution of genuine scores (orange box) is more right-handed, which results in a higher number of correct matches ($\uparrow TGR$). In contrast, impostor scores (purple box) are less left-handed, which implies high rates of false matches ($\uparrow FGR$). Genuine distributions of young males obtained the highest lower quartile (Q1) on the four systems. Young females obtain higher genuine scores than old females and males. Overall, all biometric systems perform worst on elderly males. For instance, Q1 of PC's genuine distribution of old males is the only one below $\tau$ when $FGR_{1000}$. This implies that the $FGR$ of this group is substantially higher than other groups. The distribution of impostor scores is similar across the four demographic groups and biometric systems. Impostor distributions are narrower than genuine distributions. However, the number of outliers is considerable (see LBP's distributions). Setting $\tau$ at $\sim ZFIR$, young males obtain higher $TIR$ than females and elderly adults.

\section{Politics beyond fairness in biometric systems}

Biometric systems will always be biased as we have previously demonstrated. Yet these systems are getting more accurate and fairer. The percentage of errors and biases are decreasing consistently over the years. Private companies and research centers are training their systems with better image quality and more sophisticated algorithmic architectures which outperforms previous versions. For instance, the latest publication released by NIST~\cite{nist2021fr} reported that the best vendor's facial recognition has negligible errors with VISA Border Photos ($FGR_{10^6} = 0.0023$). Moreover, the analysis on demographic disparities shows no significant differences. In addition to demystifying myths on the technical details of algorithmic systems, we argue that a critical approach to fairness in biometrics, coined as ‘critical biometric consciousness’ by Simone Browne~\citep[p.~116]{browne2015dark}, should also shift the attention towards the political context and racialised mechanism in where biometrics are embedded~\cite{amoore2021deep, castelvecchi2020beating}.

Since 2001, the EU has established a digital border infrastructure for migration control~\cite{broeders2007new, jones2019picum}. The European Asylum Dactyloscopy Database (EURODAC), the Visa Information System (VIS), the Schengen Information System (SIS II), and the new Entry-Exity System (EES) are the four main databases that EU countries use to determine responsibility for examining an asylum application, register visa applications or border-crossings. These systems are provided with biometric systems in order to register, identify and criminalise migrants~\cite{tazzioli2019making, stenum2017body, Metcalfe_Dencik_2019}. Through these databases, migrants who have entered an EU country are identified through their fingerprints. The storage of biometric samples in these databases are used to identify asylum seekers in countries which ‘are not responsible’ for their asylum petition (EURODAC) or detect people those who visa has expired (EES and VIS). However, biometric traces provided by these biometric systems are  used in other contexts such as asylum tribunal decisions. On May 2019~\footnote{Note this date is before Brexit, so the UK had fully access to EURODAC.}, the First-tier Tribunal in the UK refused a petition of asylum to an individual national of Iraq of Kurdish origin due to a biometric evidence given by one of these systems~\cite{tribunal2019},  alongside other discrepancies. The asylum seeker claimed that he was at risk of serious harm, stating that one family member had been killed and his own house was intentionally set on
fire. He appealed against the decision, the Deputy Upper Tribunal Judge did not set aside the previous decision:

\begin{quote}
A further document was relied upon by the Respondent at that hearing, being a EURODAC search result, demonstrating that a person in the Appellant's identity was fingerprinted in Dresden in Germany on 22 March 2016. The Appellant's account as given in his Statement of Evidence (SEF) interview and confirmed in oral evidence before the judge was that he only left Iraq in December 2017 The Appellant denied before the judge that the person identified in the EURODAC search was him but the judge stated that she was satisfied by the details contained within the document, and looking at the clear photograph on the EURODAC match, that the person fingerprinted in Germany was indeed the Appellant. [...] 
The evidence provided by the EURODAC document is unequivocal. The photograph contained within the document is clearly the Appellant who appeared before me and I have no
reason to doubt that the document relates to him. Therefore this document upon which I am satisfied I can place significant weight puts him in Germany on 22 March 2016.
Consequently, I find that this evidence undermines the credibility of his entire account and his credibility as a witness in his own cause and that it renders his entire account unreliable.~\citep[p.~2]{tribunal2019}
\end{quote}

This case unveils how biometric systems are used nowadays at the border to refuse and fail asylum seekers. As Browne has argued biometrics are a technology ‘to make the mute body disclose the truth of its racial identities’ and ‘that can be employed to do the work of alienating the subject by producing a truth about the racial body and one's identity (or identities) despite the subject's claims’~\citep[pp.~108-110]{browne2015dark}. The evidence given by EURODAC's fingerprint system was placed ahead of the person's narrative within the hierarchy of truthfulness~\cite{aradau2022error}. The asylum seeker was portrayed as a deceptive subject given that the biometric system unveiled the ‘truthful’ of his whereabouts. The inconsistency encountered did not question the credibility of the technology, but rather affected the asylum seeker's credibility. Moreover, this inconsistency became decisive for the UK's judicial power to refuse his asylum petition. 

Perhaps less intuitively, this asylum appeal also exposes the \textit{the elephant in the room} of biometrics. While engineers are centering their efforts in training better performing, fairer, and more equitable biometrics, the same systems are implemented at the border to deny asylum and push migrants back. As previously shown, we are witnessing a trend on fairness in biometrics~\cite{drozdowski2020demographic}. Furthermore, we also observe this trend in algorithmic fairness for migration and asylum contexts, proposing approaches to better distribute asylum seekers within a country~\cite{bansak2021algorithmic, ahmad2020refugees, kinchin2021technology}. However, algorithmic fairness cannot distribute justice in scenarios which intended purpose is to discriminate and that consistently jeopardise fundamental rights. As Tenday Achiume has argued: ‘[T]here can be no technological solution to the inequities of digital racial borders’ \citep[p.~337]{achiume2021digital}. Fairer biometrics and algorithmic solutions implemented at racialised borders conceals the injustices that these infrastructures reproduce. These social, political and historical injustices are \textit{the elephant in the room} of biometrics, the controversial issue that is obvious but remains ignored and unmentioned in debates around borders, biometrics and fairness. 

In April 2021, the European Parliament published the AI Act, the first legal framework proposal to regulate artificial intelligence~\cite{EC2021AIAAct}. The scope of the AI Act is to address the risks associated with the use of such a technology and protect safety and fundamental rights. Within this document, biometrics is considered a high-risk system in the following areas: (i) biometrics identification and categorisation of natural persons, (ii) migration, asylum and border control management, (iii) law enforcement and (iv) emotion recognition. Probably, the recent campaigns against mass surveillance using facial recognition organised by several organisations have played a key role for the regulation of biometric system~\cite{bigbrotherwatch2020fr,edri2020fr}. Indeed, the proposal opts to ban the use of ‘real-time’ facial recognition in public spaces~\cite{veale2021demystifying}. However, certain exceptions are considered regarding the use of biometrics systems such as targeted search for specific potential victims of crime,  threat to the life or physical safety of natural persons or of a terrorist attack or perpetrator or suspect of a criminal offence. Interestingly, the exception announced in Article 83 has gone completely unnoticed in the public debate on the AI Act:

\begin{quote}
This Regulation shall not apply to the AI systems which are components of the large-scale IT systems established by the legal acts listed in Annex IX that have been placed on the market or put into service before [12 months after the date of application of this Regulation referred to in Article 85(2)], unless the replacement or amendment of those legal acts leads to a significant change in the design or intended purpose of the AI system or AI systems concerned.~\citep[p.~88]{EC2021AIAAct}
\end{quote}

Despite of the fact that the legal document categorised as high-risk biometric systems used in the context of migration, border control management and law enforcement, the previous text exposes that this same regulation does not apply on the four biometric databases (EURODAC, VIS, SIS II, and EES) which are used specifically for these purposes. Regulation will only apply when there is a ‘significant change’ within these systems, but the proposed document does not provide what does a ‘significant change’ mean. Thus, the AI Act will not entail any substantial legal change for migrants and asylum seekers who are screened by biometric systems upon arrival in Europe.

Whilst facial recognition technologies used by police authorities in public spaces has been banned by the European Parliament~\cite{ojamo2021ep, sanchez2021euobserver}, the use of fingerprints to immobilise migrants will remain legalised and in the shadows of any public or political debate. As the EU Rapporteur, Petar Vitanov, said after the resolution approved by the European institution: ‘This is a huge win for all European citizens’. Yet, fundamental rights for non-Europeans will be put into the background. Asymmetries in legal, political and social rights have been historically confronted. In her classic \textit{Women, Race \& Class}, Angela Davis narrates the frictions between the (white) feminist movement and the enslavement of Black people. During the women's rights campaign in the US, she explains the advanced political position of an American abolitionist and women's rights advocate: ‘But Angelina Grimke proposed a principled defense of the unity between Black Liberation and Women's Liberation: ``I want to be identified with the Negro'', she insisted. `Until he gets his rights, we shall never have ours.''’~\citep[p.~59]{davis2011women}. Bringing Grimke's political consciousness to our discussion about the regulation of biometrics, we suggest that until migrants get their digital and fundamental rights, we shall never have ours.

\section{Discussion}

Fairness has emerged in the context of biometrics as a new research area. It aims at addressing and mitigate demographic biases in systems designed to identify or authenticate subjects based on body features (eyes, face, finger veins, among others). Yet fairness has overshadowed the \textit{the elephant in the room} of the use of biometrics, the controversial issue which is obviously present but is avoided as a subject for discussion. Biometrics has a long-standing colonial and racial legacy which is usually ignored by the biometric industry and research field. This heritage is still latent today with the implementation of biometric systems for the purposes of migration control and law enforcement. Whilst the study of fairness revolves around ‘debiasing’ biometric systems, migrants' fundamental rights are jeopardised by the use of this technology at the border.

In this paper we argued that biometrics are and will be always biased. Building on the literature of fairness in machine learning, we demonstrated theoretically that biometric systems cannot mutually satisfied different fairness definitions. Then, we empirically demonstrate the impossibility of fair biometric systems. We observed that the biometric dataset proposed to train the biometric systems reproduce racialisation of bodies, underrepreseting non-Western subjects and using race categories that are archaic and offensive. The results clearly held the theoretical framework, showing that biometric systems show differences on three fairness criteria based on age, gender and race groups at different decision thresholds. Yet, this paper has pushed this argument further showing how the focus on the fairness of biometrics undermines the political discourse about the use of biometrics at the border. As we have shown, biometric systems are used nowadays by border and judicial authorities for migration control. The algorithmic decision is used to assess the narrative of the migrant or asylum seeker, and in case of an inconsistency, the biometric output is positioned as the real truth. Moreover, the recent proposed AI regulation by the EU that bans and categorised certain biometric systems will not applied to the large-scale databases that are used to immobilise and criminalise migrants. 

In conclusion, the use of fairness in algorithmic systems installed in social and political contexts which principal and intended function is to discriminate, displaces the breach of fundamental rights because the algorithm is ‘fair’. Fairer biometric systems embedded at the border will legitimise denials of asylum, push-backs or secondary movements. Moreover, the current debates around the demographic biases and the ethics of artificial intelligence overshadows political, social and historical discrimination that was there before the technology. As Browne argues: ‘a critical biometric consciousness must acknowledge the connections between contemporary biometric information technologies and their historical antecedents’~\citep[p.~118]{browne2015dark}. As this trend will become more prominent in the incoming years, there is an urgent need to shift the debates around the colonial and racial context in which most of these systems are embedded.


\section*{Acknowledgments}
We are grateful to Martina Tazzioli for her detailed and insightful comments on early drafts. We also thank Claudia Aradau, Sarah Perret, Lucrezia Canzutti and Tobias Blanke for their expert editing advice that greatly improved the article; and Lorena Jaume-Paulasí and Reuben Binns for their ethical and data protection advice. The work of Ana Valdivia was supported by SECURITY FLOWS (ERC Consolidator Grant, grant number 819213). The work of Júlia Corbera-Serrajòrdia and Aneta Swianiewicz was supported by the King's Undergraduate Research Fellowship.

\bibliographystyle{natbib}
\bibliography{bibliography}

\begin{thebibliography}{}

\bibitem[Achiume(2021)Achiume]{achiume2021digital}
Achiume, E.~T. (2021).
\newblock Digital racial borders.
\newblock {\em American Journal of International Law\/}, {\bf 115}, 333--338.

\bibitem[Acien {\em et~al.}(2018)Acien, Morales, Vera-Rodriguez, Bartolome, and
  Fierrez]{acien2018measuring}
Acien, A., Morales, A., Vera-Rodriguez, R., Bartolome, I., and Fierrez, J.
  (2018).
\newblock Measuring the gender and ethnicity bias in deep models for face
  recognition.
\newblock In {\em Iberoamerican Congress on Pattern Recognition\/}, pages
  584--593. Springer.

\bibitem[Ahmad(2020)Ahmad]{ahmad2020refugees}
Ahmad, N. (2020).
\newblock Refugees and algorithmic humanitarianism: Applying artificial
  intelligence to {RSD} procedures and immigration decisions and making global
  human rights obligations relevant to ai governance.
\newblock {\em International Journal on Minority and Group Rights\/}, {\bf 1},
  1--69.

\bibitem[Aloudat {\em et~al.}(2016)Aloudat, Michael, and
  Abbas]{aloudat2016implications}
Aloudat, A., Michael, K., and Abbas, R. (2016).
\newblock The implications of iris-recognition technologies: Will our eyes be
  our keys?
\newblock {\em IEEE Consumer Electronics Magazine\/}, {\bf 5}(3), 95--102.

\bibitem[{Amnesty International}(2016){Amnesty International}]{amnesty2016}
{Amnesty International} (2016).
\newblock {\em Hotspot Italy: Abuses of refugees and migrants\/}.
\newblock {Amnesty International. Campaign.}
\newblock Available at:
  \url{https://www.amnesty.org/en/latest/campaigns/2016/11/hotspot-italy}
  (accessed 29 October 2021).

\bibitem[Amoore(2021)Amoore]{amoore2021deep}
Amoore, L. (2021).
\newblock The deep border.
\newblock {\em Political Geography\/}, page 102547.

\bibitem[Aradau and Perret(2022)Aradau and Perret]{aradau2022error}
Aradau, C. and Perret, S. (2022).
\newblock The politics of (non)knowledge at europe's border: Errors, fakes and
  subjectivity.
\newblock {\em Forthcoming\/}.

\bibitem[Bansak and Mart{\'e}n(2021)Bansak and
  Mart{\'e}n]{bansak2021algorithmic}
Bansak, K. and Mart{\'e}n, L. (2021).
\newblock Algorithmic decision-making, fairness, and the distribution of
  impact: Application to refugee matching in sweden.

\bibitem[Barocas {\em et~al.}(2019)Barocas, Hardt, and
  Narayanan]{barocas-hardt-narayanan}
Barocas, S., Hardt, M., and Narayanan, A. (2019).
\newblock {\em Fairness and Machine Learning\/}.
\newblock fairmlbook.org.
\newblock \url{http://www.fairmlbook.org}.

\bibitem[Benjamin(2019)Benjamin]{benjamin2019race}
Benjamin, R. (2019).
\newblock {\em Race After Technology: Abolitionist Tools for the New Jim
  Code\/}.
\newblock Cambridge, UK: Polity Press.

\bibitem[Birhane(2021)Birhane]{birhane2021impossibility}
Birhane, A. (2021).
\newblock The impossibility of automating ambiguity.
\newblock {\em Artificial Life\/}, {\bf 27}(1), 44--61.

\bibitem[Broeders(2007)Broeders]{broeders2007new}
Broeders, D. (2007).
\newblock The new digital borders of europe: Eu databases and the surveillance
  of irregular migrants.
\newblock {\em International sociology\/}, {\bf 22}(1), 71--92.

\bibitem[Browne(2015)Browne]{browne2015dark}
Browne, S. (2015).
\newblock {\em Dark matters\/}.
\newblock Durham and London: Duke University Press.

\bibitem[Buolamwini and Gebru(2018)Buolamwini and Gebru]{buolamwini2018gender}
Buolamwini, J. and Gebru, T. (2018).
\newblock Gender shades: Intersectional accuracy disparities in commercial
  gender classification.
\newblock In {\em Conference on fairness, accountability and transparency\/},
  pages 77--91. PMLR.

\bibitem[Butler(1999)Butler]{butler1999gender}
Butler, J. (1999).
\newblock {\em Gender Trouble\/}.
\newblock London: Routledge.

\bibitem[Castelvecchi(2020)Castelvecchi]{castelvecchi2020beating}
Castelvecchi, D. (2020).
\newblock Beating biometric bias.
\newblock {\em Nature\/}, {\bf 587}(7834), 347--349.

\bibitem[Choi {\em et~al.}(2009)Choi, Song, Kim, Lee, and Kim]{choi2009finger}
Choi, J.~H., Song, W., Kim, T., Lee, S.-R., and Kim, H.~C. (2009).
\newblock Finger vein extraction using gradient normalization and principal
  curvature.
\newblock In {\em Image Processing: Machine Vision Applications II\/}, volume
  7251, page 725111. International Society for Optics and Photonics.

\bibitem[Chouldechova(2017)Chouldechova]{chouldechova2017fair}
Chouldechova, A. (2017).
\newblock Fair prediction with disparate impact: A study of bias in recidivism
  prediction instruments.
\newblock {\em Big data\/}, {\bf 5}(2), 153--163.

\bibitem[Davis(2019)Davis]{davis2011women}
Davis, A.~Y. (2019).
\newblock {\em Women, Race, \& Class\/}.
\newblock London: Penguin Random House UK.

\bibitem[de~Freitas~Pereira and Marcel(2020)de~Freitas~Pereira and
  Marcel]{de2020fairness}
de~Freitas~Pereira, T. and Marcel, S. (2020).
\newblock Fairness in biometrics: a figure of merit to assess biometric
  verification systems.
\newblock {\em arXiv preprint arXiv:2011.02395\/}.

\bibitem[Drozdowski {\em et~al.}(2020)Drozdowski, Rathgeb, Dantcheva, Damer,
  and Busch]{drozdowski2020demographic}
Drozdowski, P., Rathgeb, C., Dantcheva, A., Damer, N., and Busch, C. (2020).
\newblock Demographic bias in biometrics: A survey on an emerging challenge.
\newblock {\em IEEE Transactions on Technology and Society\/}, {\bf 1}(2),
  89--103.

\bibitem[Drozdowski {\em et~al.}(2021)Drozdowski, Prommegger, Wimmer, Schraml,
  Rathgeb, Uhl, and Busch]{drozdowski2021demographic}
Drozdowski, P., Prommegger, B., Wimmer, G., Schraml, R., Rathgeb, C., Uhl, A.,
  and Busch, C. (2021).
\newblock Demographic bias: A challenge for fingervein recognition systems?
\newblock In {\em 2020 28th European Signal Processing Conference (EUSIPCO)\/},
  pages 825--829. IEEE.

\bibitem[Dunkelau and Leuschel(2019)Dunkelau and
  Leuschel]{dunkelau2019fairness}
Dunkelau, J. and Leuschel, M. (2019).
\newblock Fairness-aware machine learning.

\bibitem[Dwork {\em et~al.}(2012)Dwork, Hardt, Pitassi, Reingold, and
  Zemel]{dwork2012fairness}
Dwork, C., Hardt, M., Pitassi, T., Reingold, O., and Zemel, R. (2012).
\newblock Fairness through awareness.
\newblock In {\em Proceedings of the 3rd innovations in theoretical computer
  science conference\/}, pages 214--226.

\bibitem[{EDRi}(2020){EDRi}]{edri2020fr}
{EDRi} (2020).
\newblock {\em Facial Recognition \& Biometric Mass Surveillance: Document
  Pool\/}.
\newblock EDRi. Report, Brussels, Belgium, March.
\newblock Available at:
  \url{https://edri.org/our-work/facial-recognition-document-pool/} (accessed
  29 October 2021).

\bibitem[{European Commission}(2021){European Commission}]{EC2021AIAAct}
{European Commission} (2021).
\newblock {\em Proposal for a Regulation of the European Parliament and of the
  Council laying down harmonised rules on artificial intelligence (Artificial
  Intelligence Act) and amending certain Union legislative acts (COM(2021) 206
  final)\/}.

\bibitem[Fang {\em et~al.}(2021)Fang, Damer, Kirchbuchner, and
  Kuijper]{fang2021demographic}
Fang, M., Damer, N., Kirchbuchner, F., and Kuijper, A. (2021).
\newblock Demographic bias in presentation attack detection of iris recognition
  systems.
\newblock In {\em 2020 28th European Signal Processing Conference (EUSIPCO)\/},
  pages 835--839. IEEE.

\bibitem[Feng {\em et~al.}(2016)Feng, Chao, and Jialiang]{feng2016finger}
Feng, L., Chao, W., and Jialiang, P. (2016).
\newblock Finger vein recognition using log gabor filter and local derivative
  pattern.
\newblock {\em International Journal of Signal Processing, Image Processing and
  Pattern Recognition (IJSIP)\/}, {\bf 9}(12), 231--242.

\bibitem[Garg {\em et~al.}(2020)Garg, Villasenor, and Foggo]{garg2020fairness}
Garg, P., Villasenor, J., and Foggo, V. (2020).
\newblock Fairness metrics: A comparative analysis.
\newblock In {\em 2020 IEEE International Conference on Big Data (Big Data)\/},
  pages 3662--3666. IEEE.

\bibitem[Gilroy(2000)Gilroy]{gilroy2000against}
Gilroy, P. (2000).
\newblock {\em Against race: Imagining political culture beyond the color
  line\/}.
\newblock Cambridge, Massachusetts: Harvard University Press.

\bibitem[Glouftsios and Scheel(2021)Glouftsios and
  Scheel]{glouftsios2021inquiry}
Glouftsios, G. and Scheel, S. (2021).
\newblock An inquiry into the digitisation of border and migration management:
  performativity, contestation and heterogeneous engineering.
\newblock {\em Third World Quarterly\/}, {\bf 42}(1), 123--140.

\bibitem[Grother {\em et~al.}(2019)Grother, Ngan, and Hanaoka]{grother2019face}
Grother, P., Ngan, M., and Hanaoka, K. (2019).
\newblock {\em Face recognition vendor test (FRVT)\/}.
\newblock US Department of Commerce, National Institute of Standards and
  Technology.

\bibitem[Grother {\em et~al.}(2021)Grother, Ngan, and Kayee]{nist2021fr}
Grother, P., Ngan, M., and Kayee, H. (2021).
\newblock {\em Ongoing Face Recognition Vendor Test (FRVT) Part 1:
  Verification\/}.
\newblock Available at:
  \url{https://www.nist.gov/programs-projects/face-recognition-vendor-test-frvt-ongoing
  } (accessed 09 November 2021).

\bibitem[Hardt {\em et~al.}(2016)Hardt, Price, and Srebro]{hardt2016equality}
Hardt, M., Price, E., and Srebro, N. (2016).
\newblock Equality of opportunity in supervised learning.
\newblock {\em Advances in neural information processing systems\/}, {\bf 29},
  3315--3323.

\bibitem[Hutchinson and Mitchell(2019)Hutchinson and
  Mitchell]{hutchinson2019unfairness}
Hutchinson, B. and Mitchell, M. (2019).
\newblock 50 years of test (un)fairness: Lessons for machine learning.
\newblock In {\em Proceedings of the Conference on Fairness, Accountability,
  and Transparency\/}, pages 49--58.

\bibitem[(Immigration and Chamber)(2019)(Immigration and
  Chamber)]{tribunal2019}
(Immigration, U.~T. and Chamber), A. (2019).
\newblock {\em Appeal Number: PA/00240/2019\/}.
\newblock Available at:
  \url{https://tribunalsdecisions.service.gov.uk/utiac/pa-00240-2019 }
  (accessed 01 November 2021).

\bibitem[Jones(2019)Jones]{jones2019picum}
Jones, C. (2019).
\newblock Data protection, immigration, enforcement and fundamental rights.
\newblock {\em Statewatch\/}.

\bibitem[Kantayya(2020)Kantayya]{kantayya2020coded}
Kantayya, S. (2020).
\newblock Coded bias.

\bibitem[Kauba {\em et~al.}(2014)Kauba, Reissig, and Uhl]{kauba2014pre}
Kauba, C., Reissig, J., and Uhl, A. (2014).
\newblock Pre-processing cascades and fusion in finger vein recognition.
\newblock In {\em 2014 International Conference of the Biometrics Special
  Interest Group (BIOSIG)\/}, pages 1--6. IEEE.

\bibitem[Kauba {\em et~al.}(2018)Kauba, Prommegger, and
  Uhl]{kauba2018focussing}
Kauba, C., Prommegger, B., and Uhl, A. (2018).
\newblock Focussing the beam-a new laser illumination based data set providing
  insights to finger-vein recognition.
\newblock In {\em 2018 IEEE 9th International Conference on Biometrics Theory,
  Applications and Systems (BTAS)\/}, pages 1--9. IEEE.

\bibitem[Kinchin(2021)Kinchin]{kinchin2021technology}
Kinchin, N. (2021).
\newblock Technology, displaced? the risks and potential of artificial
  intelligence for fair, effective, and efficient refugee status determination.
\newblock {\em Law in Context. A Socio-legal Journal\/}, {\bf 37}(3).

\bibitem[Kleinberg(2018)Kleinberg]{kleinberg2018inherent}
Kleinberg, J. (2018).
\newblock Inherent trade-offs in algorithmic fairness.
\newblock In {\em Abstracts of the 2018 ACM International Conference on
  Measurement and Modeling of Computer Systems\/}, pages 40--40.

\bibitem[Lohr(2018)Lohr]{lohr2018facial}
Lohr, S. (2018).
\newblock Facial recognition is accurate, if you’re a white guy.
\newblock {\em New York Times\/}, {\bf 9}(8), 283.

\bibitem[Lu {\em et~al.}(2013)Lu, Xie, Yoon, Wang, and Park]{lu2013available}
Lu, Y., Xie, S.~J., Yoon, S., Wang, Z., and Park, D.~S. (2013).
\newblock An available database for the research of finger vein recognition.
\newblock In {\em 2013 6th International congress on image and signal
  processing (CISP)\/}, volume~1, pages 410--415. IEEE.

\bibitem[Maguire(2009)Maguire]{maguire2009birth}
Maguire, M. (2009).
\newblock The birth of biometric security.
\newblock {\em Anthropology today\/}, {\bf 25}(2), 9--14.

\bibitem[Marasco(2019)Marasco]{marasco2019biases}
Marasco, E. (2019).
\newblock Biases in fingerprint recognition systems: Where are we at?
\newblock In {\em 2019 IEEE 10th International Conference on Biometrics Theory,
  Applications and Systems (BTAS)\/}, pages 1--5. IEEE.

\bibitem[Metcalfe and Dencik(2019)Metcalfe and Dencik]{Metcalfe_Dencik_2019}
Metcalfe, P. and Dencik, L. (2019).
\newblock The politics of big borders: Data (in)justice and the governance of
  refugees.
\newblock {\em First Monday\/}, {\bf 24}(4).

\bibitem[Miura {\em et~al.}(2007)Miura, Nagasaka, and
  Miyatake]{miura2007extraction}
Miura, N., Nagasaka, A., and Miyatake, T. (2007).
\newblock Extraction of finger-vein patterns using maximum curvature points in
  image profiles.
\newblock {\em IEICE TRANSACTIONS on Information and Systems\/}, {\bf 90}(8),
  1185--1194.

\bibitem[O'Flaherty(2020)O'Flaherty]{o2020facial}
O'Flaherty, M. (2020).
\newblock Facial recognition technology and fundamental rights.
\newblock {\em Eur. Data Prot. L. Rev.}, {\bf 6}, 170.

\bibitem[Ojamo(2021)Ojamo]{ojamo2021ep}
Ojamo, J. (2021).
\newblock {\em Use of artificial intelligence by the police: MEPs oppose mass
  surveillance\/}.
\newblock European Parliament.
\newblock Available at:
  \url{https://www.europarl.europa.eu/news/en/press-room/20210930IPR13925/use-of-artificial-intelligence-by-the-police-meps-oppose-mass-surveillance}
  (accessed 12 November 2021).

\bibitem[Preciozzi {\em et~al.}(2020)Preciozzi, Garella, Camacho, Franzoni,
  Di~Martino, Carbajal, and Fernandez]{preciozzi2020fingerprint}
Preciozzi, J., Garella, G., Camacho, V., Franzoni, F., Di~Martino, L.,
  Carbajal, G., and Fernandez, A. (2020).
\newblock Fingerprint biometrics from newborn to adult: A study from a national
  identity database system.
\newblock {\em IEEE Transactions on Biometrics, Behavior, and Identity
  Science\/}, {\bf 2}(1), 68--79.

\bibitem[Queiroz(2019)Queiroz]{queiroz2019impact}
Queiroz, B.~M. (2019).
\newblock The impact of eurodac in eu migration law: The era of crimmigration?
\newblock {\em Market and Competition Law Review\/}, {\bf 3}(1), 157--183.

\bibitem[Ross {\em et~al.}(2019)Ross, Banerjee, Chen, Chowdhury, Mirjalili,
  Sharma, Swearingen, and Yadav]{ross2019some}
Ross, A., Banerjee, S., Chen, C., Chowdhury, A., Mirjalili, V., Sharma, R.,
  Swearingen, T., and Yadav, S. (2019).
\newblock Some research problems in biometrics: The future beckons.
\newblock In {\em 2019 International Conference on Biometrics (ICB)\/}, pages
  1--8. IEEE.

\bibitem[Scheel(2013)Scheel]{scheel2013autonomy}
Scheel, S. (2013).
\newblock Autonomy of migration despite its securitisation? facing the terms
  and conditions of biometric rebordering.
\newblock {\em Millennium\/}, {\bf 41}(3), 575--600.

\bibitem[Serna {\em et~al.}(2021)Serna, Pe{\~n}a, Morales, and
  Fierrez]{serna2021insidebias}
Serna, I., Pe{\~n}a, A., Morales, A., and Fierrez, J. (2021).
\newblock Insidebias: Measuring bias in deep networks and application to face
  gender biometrics.
\newblock In {\em 2020 25th International Conference on Pattern Recognition
  (ICPR)\/}, pages 3720--3727. IEEE.

\bibitem[Stenum(2017)Stenum]{stenum2017body}
Stenum, H. (2017).
\newblock The body-border. governing irregular migration through biometric
  technology.
\newblock {\em spheres: Journal for Digital Cultures\/}, {\bf 4}, 1--16.

\bibitem[Sánchez~Nicolás(2021)Sánchez~Nicolás]{sanchez2021euobserver}
Sánchez~Nicolás, E. (2021).
\newblock {\em MEPs back EU facial-recognition ban for police\/}.
\newblock EUobserver.
\newblock Available at: \url{https://euobserver.com/democracy/153135} (accessed
  12 November 2021).

\bibitem[Tazzioli(2019)Tazzioli]{tazzioli2019making}
Tazzioli, M. (2019).
\newblock {\em The making of migration: The biopolitics of mobility at
  Europe’s borders\/}.
\newblock Thousand Oaks, California: SAGE.

\bibitem[Terh{\"o}rst {\em et~al.}(2021)Terh{\"o}rst, Kolf, Huber,
  Kirchbuchner, Damer, Morales, Fierrez, and
  Kuijper]{terhorst2021comprehensive}
Terh{\"o}rst, P., Kolf, J.~N., Huber, M., Kirchbuchner, F., Damer, N., Morales,
  A., Fierrez, J., and Kuijper, A. (2021).
\newblock A comprehensive study on face recognition biases beyond demographics.
\newblock {\em arXiv preprint arXiv:2103.01592\/}.

\bibitem[Ton and Veldhuis(2013)Ton and Veldhuis]{ton2013high}
Ton, B.~T. and Veldhuis, R.~N. (2013).
\newblock A high quality finger vascular pattern dataset collected using a
  custom designed capturing device.
\newblock In {\em 2013 International conference on biometrics (ICB)\/}, pages
  1--5. IEEE.

\bibitem[Uhl {\em et~al.}(2020)Uhl, Busch, Marcel, and
  Veldhuis]{uhl2020handbook}
Uhl, A., Busch, C., Marcel, S., and Veldhuis, R. (2020).
\newblock {\em Handbook of vascular biometrics\/}.
\newblock Springer Nature.

\bibitem[Van~der Ploeg(1999)Van~der Ploeg]{van1999illegal}
Van~der Ploeg, I. (1999).
\newblock The illegal body:eurodac'and the politics of biometric
  identification.
\newblock {\em Ethics and Information Technology\/}, {\bf 1}(4), 295--302.

\bibitem[Vanoni {\em et~al.}(2014)Vanoni, Tome, El~Shafey, and
  Marcel]{vanoni2014cross}
Vanoni, M., Tome, P., El~Shafey, L., and Marcel, S. (2014).
\newblock Cross-database evaluation using an open finger vein sensor.
\newblock In {\em 2014 IEEE workshop on biometric measurements and systems for
  security and medical applications (BIOMS) proceedings\/}, pages 30--35. IEEE.

\bibitem[Veale and Borgesius(2021)Veale and Borgesius]{veale2021demystifying}
Veale, M. and Borgesius, F.~Z. (2021).
\newblock Demystifying the draft eu artificial intelligence act—analysing the
  good, the bad, and the unclear elements of the proposed approach.
\newblock {\em Computer Law Review International\/}, {\bf 22}(4), 97--112.

\bibitem[Verma and Rubin(2018)Verma and Rubin]{verma2018fairness}
Verma, S. and Rubin, J. (2018).
\newblock Fairness definitions explained.
\newblock In {\em 2018 ieee/acm international workshop on software fairness
  (fairware)\/}, pages 1--7. IEEE.

\bibitem[Watch(2020)Watch]{bigbrotherwatch2020fr}
Watch, B.~B. (2020).
\newblock {\em Big Brother Watch Briefing on facial recognition
  surveillance\/}.
\newblock Available at:
  \url{https://bigbrotherwatch.org.uk/wp-content/uploads/2020/06/Big-Brother-Watch-briefing-on-Facial-recognition-surveillance-June-2020.pdf}
  (accessed 29 October 2021).

\bibitem[Xie and Kumar(2019)Xie and Kumar]{xie2019finger}
Xie, C. and Kumar, A. (2019).
\newblock Finger vein identification using convolutional neural network and
  supervised discrete hashing.
\newblock {\em Pattern Recognition Letters\/}, {\bf 119}, 148--156.

\bibitem[Zhao and Gordon(2019)Zhao and Gordon]{zhao2019inherent}
Zhao, H. and Gordon, G. (2019).
\newblock Inherent tradeoffs in learning fair representations.
\newblock {\em Advances in neural information processing systems\/}, {\bf 32},
  15675--15685.

\end{thebibliography}


\end{document}